\documentclass[aps,pre,twocolumn,superscriptaddress]{revtex4}
\usepackage{graphicx,amssymb,color}

\begin{document}

% Use the \preprint command to place your local institutional report
% number in the upper righthand corner of the title page in preprint mode.
% Multiple \preprint commands are allowed.
% Use the 'preprintnumbers' class option to override journal defaults
% to display numbers if necessary
%\preprint{}

%Title of paper
\title{Heterogeneity and chaos in the Peyrard-Bishop-Dauxois DNA model}

\author{M.~Hillebrand} \email{malcolm.hillebrand@gmail.com} \affiliation{Department of Mathematics and
Applied Mathematics, University of Cape Town, Rondebosch 7701,
South Africa}
\author{G.~Kalosakas}  \email{georgek@upatras.gr} \affiliation{Department of Materials Science, University of Patras, GR-26504 Rio, Greece}
\author{A.~Schwellnus} \email{SCHADR008@myuct.ac.za} \affiliation{Department of Mathematics and
Applied Mathematics, University of Cape Town, Rondebosch 7701,
South Africa}
\author{Ch.~Skokos} \email{haris.skokos@uct.ac.za}
\homepage{http://math\_research.uct.ac.za/~hskokos/}
\affiliation{Department of Mathematics and
Applied Mathematics, University of Cape Town, Rondebosch 7701,
South Africa}
\affiliation{Max Planck Institute for the Physics of Complex Systems, N\"othnitzer Str.~38, D-01187 Dresden, Germany}

\date{\today}

\begin{abstract}
% insert abstract here
We discuss the effect of heterogeneity on the chaotic properties of the Peyrard-Bishop-Dauxois nonlinear
model of DNA. Results are presented for the maximum Lyapunov exponent and the deviation vector distribution.
Different compositions of adenine-thymine (AT) and guanine-cytosine (GC) base pairs are examined for various energies up to the melting point
of the corresponding sequence. We also consider the effect of the alternation index, which measures the
heterogeneity of the DNA chain through the number of alternations between different types (AT or GC)
of base pairs, on the chaotic behavior of the system.
Biological gene promoter sequences have been also investigated, showing no distinct behavior of the
maximum Lyapunov exponent.
\end{abstract}

% insert suggested PACS numbers in braces on next line
% \pacs{05.45.-a, ????}
% insert suggested keywords - APS authors don't need to do this
% \keywords{}

%\maketitle must follow title, authors, abstract, \pacs, and \keywords
\maketitle

%_______________________________________________
\section{Introduction\label{sec:intro}}

There exists a large number of models devoted to studying the dynamics of base pairs in DNA
(see for example \cite{physrep} and references therein). Among them, the Peyrard-Bishop-Dauxois (PBD)
model \cite{PBD} has been extensively used to account for a number of experimental observations related to
base pair openings in DNA. These include the denaturation transition of short oligonucleotides \cite{CG}
and of peculiar periodic sequences \cite{boianNAR9}, the multi-step melting of heterogeneous DNA segments \cite{CH},
the formation of bubbles in the premelting regime \cite{aresPRL},
as well as large openings due to thermal fluctuations in gene promoters of various organisms at positions
related to transcriptionally relevant sites \cite{NAR,EPL,raptiBJ,boianPLOSCB,boianNAR2,angeliki,faloPRE12,huangJBE,faloPLOS}.

The PBD model is an one-dimensional coarse-grained lattice model at the base pair level, considering a continuous
variable at each site which describes the stretching of individual base pairs along the DNA sequence.
The model is an extension of an earlier version \cite{PB,DPB93,peyrardRev}, where a nonlinear stacking interaction term
has been incorporated to mimic entropic effects, resulting in a sharp denaturation transition \cite{PBD,DP95}.
Exact numerical results regarding the partition function and thermodynamic functions of the PBD model
have been obtained through the transfer integral operator \cite{DP95}. Statistical distributions of bubble lengths
for various temperatures and different guanine-cytosine content of the DNA chain have been presented
using Monte Carlo simulations \cite{saul1,saul2}. Bubble length distributions and their equilibrium properties
have been also discussed in detail for homogeneous DNA chains \cite{theodPRE}.

Apart from various statistical properties, dynamical aspects of the PBD model have been also explored.
The dynamic structure factor of a bacteriophage regulatory sequence has been evaluated for a range
of temperatures and particular features attributed to localized thermal openings were observed \cite{VKRB}.
It has been further shown that the model exhibits a complex temporal decay of the local displacement
or energy autocorrelation functions in a wide temperature range for homogeneous DNA sequences \cite{CPL};
distinct decay processes are obtained at subpicoseconds and in the picoseconds to nanoseconds time scales.
More recently, protein aggregation and oligomerization has been studied through the coalescence of protein-induced
DNA bubbles evolved according to the PBD model \cite{voulgSR}.

Concerning intrinsic localized modes, a detailed study on the earlier version of the model has led to estimates of their
characteristic size and lifetime in a thermalized homogeneous system \cite{farago}. Sub-harmonic discrete breathers
have been discussed in the driven PBD model, arising from the anti-continuous limit, even for driving frequencies
above the linear frequency of the Morse on-site potential where usual breather solutions do not exist \cite{maniadisPRE}.
In another context, the spontaneous formation of vibrational hot spots in homogeneous PBD lattices \cite{JCP03}
affects macroscopic transport parameters of a charge carrier coupled to DNA structural dynamics \cite{PRE05,PRE11},
while, in reverse, electric current is able to form bubbles by exciting the base pairs \cite{gufu}.

There are not many studies regarding calculations of the Lyapunov exponents or other indicators of chaos
in the PBD model. An early work investigated the behavior of the maximum Lyapunov exponent (mLE) of a homogeneous DNA sequence
\cite{barreEPL}. Though the focus of that work was the former version of the model, with a linear stacking interaction
and a smoother denaturation transition, where accurate analytical estimates were presented for the mLE through a combination of a Riemannian geometry approach with the transfer integral operator method,
the dependence of the  mLE on the energy density was also presented for the PBD model.
Moreover it was pointed out by the authors that the mLE, showing an abrupt change at the critical point,
could serve as a dynamical order parameter indicating a phase transition.
To the best of our knowledge there exist no investigations of the effects of the sequence heterogeneity, which is always
present in actual DNA molecules, on the chaotic behavior of the PBD model. This is the subject of the present work.

The paper is organized as follows. In Sect.~\ref{sec:model} we outline the PBD model and lay out the numerical techniques we use in our study, presenting also the quantities  we  examine. Sect.~\ref{sec:results} contains the numerical  results of our investigations and a discussion thereof. Finally, in Sect.~\ref{sec:sum} we summarize our findings and present our conclusions.

%_______________________________________________
\section{DNA model and numerical techniques\label{sec:model}}

The Hamiltonian function of the PBD model, which describes the displacements from equilibrium $y_i$ of the bases forming the $i$th base pair in a DNA sequence of $n$ base pairs, is given by \cite{PBD}
\begin{eqnarray}
      H = & \sum_{i = 1}^{n} \left[ \frac{1}{2m}p_{i}^2 + D_i(e^{-a_iy_i} - 1)^2  \right] +  \nonumber \\
       &  \sum_{i=2}^{n} \left[ \frac{K}{2}(1 + \rho e^{-b(y_{i} +y_{i-1} )})(y_i - y_{i-1})^2 \right].
\label{eq:hamiltonian}
\end{eqnarray}
The Hamiltonian consists of three terms: the kinetic energy part with $p_i$ denoting the conjugate momentum of $y_i$, a Morse potential to model the effective interaction energy of the complementary bases within the base pair at the $i$th site, and an anharmonic coupling term between first neighbors to account for the effect of stacking interaction.

To model the inhomogeneous nature of an actual DNA chain, different parameters are used in the Morse potential for adenine-thymine (AT) and guanine-cytosine (GC) base pairs. As the first sum in Hamiltonian (\ref{eq:hamiltonian}) runs over each base pair the use of different parameters for each pair  represents the  disordered behavior of the DNA sequence. The parameter values we use in our study  are $m=300$ amu for the effective mass of base pairs, for the Morse potential we have $D_{GC} = 0.075$ eV, $a_{GC} = 6.9$ \AA $^{-1}$ for GC base pairs and $D_{AT} = 0.05$ eV, $a_{AT} = 4.2$ \AA$^{-1}$ for AT base pairs, while for the stacking interaction we set $K = 0.025$ eV/\AA$^{-2}$, $\rho = 2$, and $b = 0.35$ \AA$^{-1}$. These parameters were fitted in \cite{CG} to accurately model the melting curves of short DNA sequences and subsequently used in a number of studies.

The PBD model is numerically integrated using the fourth order symplectic Runge-Kutta-Nystr\"om method \cite{blanesrk}. Symplectic integrators are a class of numerical integration methods devised particularly for Hamiltonian systems. One of their main advantages is the ability to  accurately integrate Hamiltonian systems keeping their energy $E$ (i.e.~the value of their Hamiltonian function)  bounded  for very long times (see e.g.~Chapt.~VI of \cite{hairergeom} and references therein). In our simulations the time unit is set to 1 ps and an integration time step $\tau=0.011$ ps kept the relative energy error $|E(t) - E(0)|/E(0)$ smaller than $10^{-6}$.

In all our simulations the initial conditions for the position coordinates are at equilibrium, i.e.~$y_i=0$, $i=1,2,\ldots, n$ and the momentum coordinates $p_i$, $i=1,2,\ldots, n$ are chosen randomly from a standard normal distribution with a zero mean and
unit variance and then scaled to obtain the desired total energy $E$ (or equivalently the energy density  $E_n = E/n$).
Lattices of $n=100$ sites are considered for all simulations (apart from the cases of biological promoters, see below)
and periodic boundary conditions are imposed, i.e.~$p_0=p_n, \ y_0=y_n$, and $p_{n+1}=p_1,\ y_{n+1}=y_1$.

In order to investigate the chaoticity of the PBD model the mLE, $\chi_1$, is  calculated by following the so-called standard method \cite{benettinetal,S_10}. The mLE can be used to discriminate between regular and chaotic motions as $\chi_1=0$ for regular orbits and $\chi_1>0$ for chaotic orbits. The magnitude of the mLE can also be used as a measure of the chaoticity: larger mLE values correspond to more chaotic behaviors. In practice we estimate $\chi_1$ by computing the finite time mLE
\begin{equation}
    \label{eq:finitemle}
    \chi = \frac{1}{t}\mathrm{ln}\frac{||\mathbf{w}(t)||}{||\mathbf{w}(0)||},
\end{equation}
where $\mathbf{w}(0)$ and $\mathbf{w}(t)$ are deviation  vectors from the studied orbit in the system's phase space at times $t=0$ and $t>0$ respectively, and $|| \cdot ||$ denotes the usual Euclidean vector norm. Then the mLE $\chi_1$ is $\chi_1 = \lim_{t\rightarrow \infty} \chi$. As can be easily seen from Eq.~(\ref{eq:finitemle}) the mLE is measured in inverse time units. Thus, in our study the $\chi$ is measured in ps$^{-1}$.

To efficiently and accurately follow the evolution of the deviation vector $\mathbf{w}(t)$  we numerically integrate the so-called variational equations \cite{contopoulos}, which govern the vector's dynamics. The Hamilton equations of motion and the variational ones are evolved alongside each other using the tangent map method outlined in \cite{skokostm,GS11,GES12}. In this way we obtain a numerical estimation of the mLE $\chi_1$ after long enough integration times, which are typically of the order of $10^{5}$ ps. The initial choice of deviation vector is a normalized vector whose random coordinates are uniformly chosen from the interval $[-1,1]$.

As a further investigation of the system's chaotic behavior, the normalized deviation vector distribution (DVD)
\begin{equation}
    \label{eq:dvd}
    \xi_i = \frac{\delta y_i^2 + \delta p_i ^2}{\sum_{i=1}^n \left(\delta y_i^2 + \delta p_i^2 \right)}
\end{equation}
is examined. Here, the $\delta y_i$ and $\delta p_i$ are the position and momentum coordinates of the deviation vector $\mathbf{w}(t)$ respectively, i.e.~$\mathbf{w} (t)  = \left(\delta y_1 (t), \ldots, \delta y_n (t),
\delta p_1 (t), \ldots, \delta p_n (t) \right)$. The DVD gives a measure of the sensitivity of a certain region of the chain to small variations of initial conditions, and provides some idea of the `strength of the nonlinearity' at each site as the system evolves \cite{skokosdvd,SMS18}.

To investigate the effect of the chain's heterogeneity on its chaoticity, two quantities are considered. First, the AT/GC composition of the chain, quantified by the percentage of AT base pairs $P_{AT}$, where $P_{AT}=0\%$ means a pure GC chain, and $P_{AT}=100\%$ means a pure AT chain. The second measure is the alternation index $\alpha$ \cite{RCDalpha}, which measures the chain's heterogeneity by counting the number of times that the base pair type alternates (from AT to GC or from GC to AT) along the DNA chain. As such, a large alternation index corresponds to a `well-mixed' chain, which in some respect may be considered quite homogeneous. A small alternation index corresponds to a `chunky' chain, considered more heterogeneous. The effects of both these measures on the chaoticity of the system are discussed in Sect.~\ref{sec:results}.

In all cases of different AT percentages and for all energies/temperatures up to melting, the major part of the
total energy (around 50\%) is contained in the Morse potential term of the Hamiltonian. For small energy densities
the kinetic energy contributes equally to this term, or slightly more as $P_{AT}$ increases, while the stacking interaction
is negligibly small. However, as the energy density increases the kinetic energy contribution drops, reaching around 30\%
of the total energy at melting temperature, in favor of both stacking and Morse potential terms. The stacking interaction
contribution increases with energy density, but even at melting stays below 20\%.

%_______________________________________________
\section{Numerical results\label{sec:results}}

%_______________________________________________
\subsection{Chaotic behavior of the PBD model \label{ssec:PAT}}

In order to investigate the chaoticity of the PBD model we calculated the dependence of the finite time mLE $\chi$ on the percentage of AT base pairs, $P_{AT}$, and on the energy density $E_n$. In particular, for each $P_{AT}$ and $E_n$ value considered we estimate the corresponding mLE $\chi_1$ by performing statistical averages over 100 different simulations. First, 10 random disorder realizations, i.e.~arrangements of AT and GC base pairs  satisfying the considered $P_{AT}$ requirement, are chosen and then, for each one of these different realizations 10 random initial conditions are evolved in time. The 100 evolutions of $\chi(t)$  created from these simulations are then used to provide an average $\langle \chi(t) \rangle$ value. In cases where there are not 10 distinct realizations (i.e. for the homogeneous cases of pure AT and pure GC, where $P_{AT}=100\%$ and $P_{AT}=0\%$ respectively) then 100 random initial conditions are considered.

%%%%%%%%%%%%%%%%%%%%%%%%%%%%%%%%
\begin{figure}
    \includegraphics[width=0.5\textwidth]{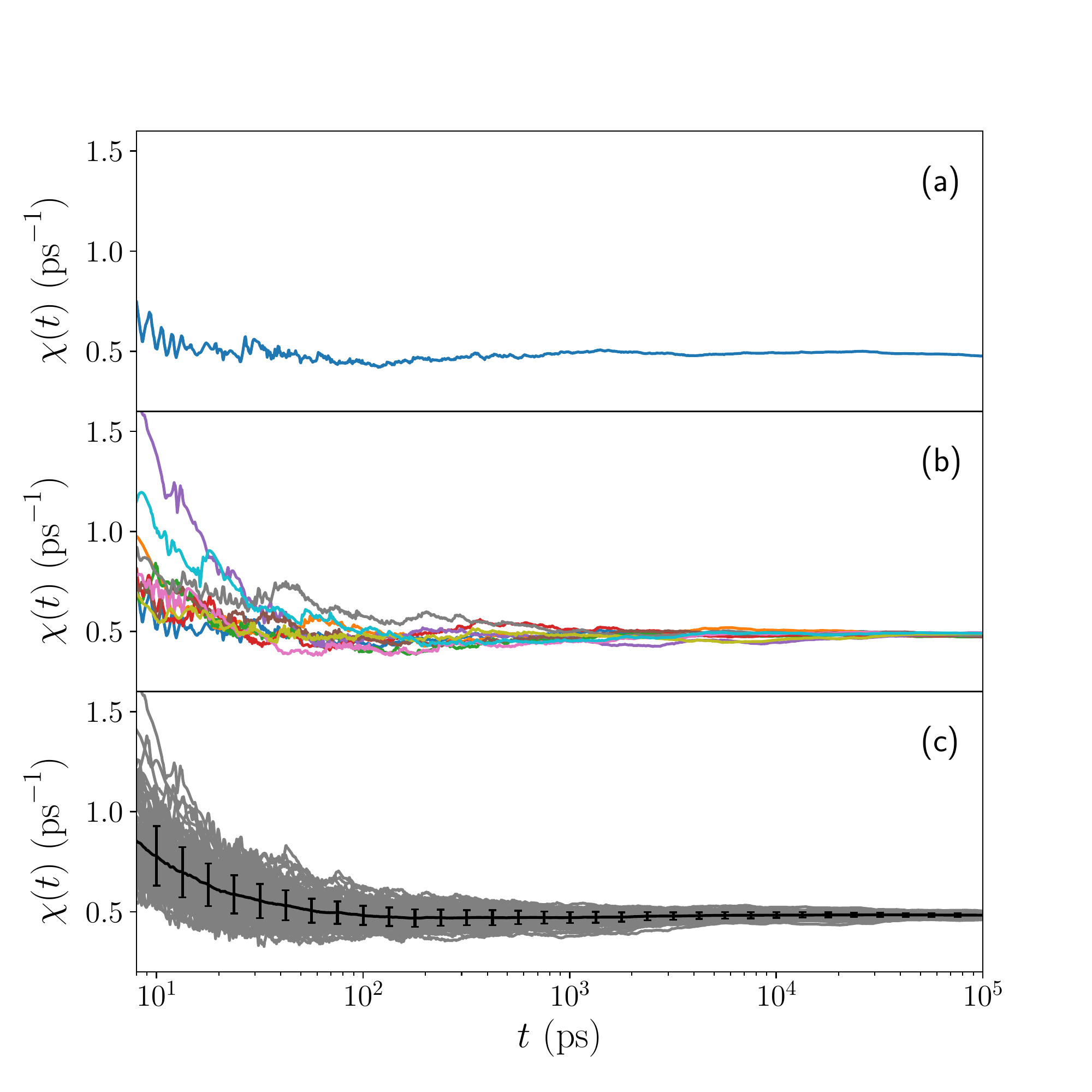}
    \caption{\label{fig:mlebuild} An illustration of the calculation of an average finite time mLE $\langle \chi \rangle$ for the particular case of  $P_{AT}=30\%$ at a given energy density $E_n=0.0475$ eV. The time evolution of $\chi$ is shown for (a)  a single initial condition and disorder realization, (b) 10 different initial conditions over the same disorder realization, and (c) 100 simulations obtained from 10 random initial conditions of each one of 10 different disorder realizations (gray curves). The computed averaged finite time mLE $\langle \chi \rangle$ along with its standard deviation (error bars) are plotted in black in (c).}
\end{figure}
%%%%%%%%%%%%%%%%%%%%%%%%%%%%%%%%

The process mentioned above for the calculation of $\langle \chi(t) \rangle$ is illustrated in Fig.~\ref{fig:mlebuild},
where the particular case of $P_{AT}=30\%$ and $E_n=0.0475$ eV is presented. In Fig.~\ref{fig:mlebuild}(a) the time evolution
of $\chi$ for a particular disorder realization compatible with the value $P_{AT}=30\%$ and an initial condition satisfying $E_n=0.0475$ eV is shown. It is seen that after some initial fluctuations $\chi(t)$  shows the tendency to stabilize at a positive value
$\chi \approx 0.47$ ps$^{-1}$. In Fig.~\ref{fig:mlebuild}(b) the same tendency is seen for the evolution of $\chi$ for 10 different initial
conditions (all of which set $E_n=0.0475$ eV) for the same disorder realization as that of Fig.~\ref{fig:mlebuild}(a).
This behavior clearly indicates that the limiting value of $\chi$ does not dependent on the particular  initial condition.
Furthermore, this value seems to not depend much on the particular random disorder realization as  the computed finite time mLEs
for all the 100 initial conditions considered for the 10 different disorder realizations in the  $P_{AT}=30\%$, $E_n=0.0475$ eV
case [gray curves in Fig.~\ref{fig:mlebuild}(c)] tend to roughly the same positive value, exhibiting only relatively small differences.

The results of Fig.~\ref{fig:mlebuild} indicate that the chaotic behavior of the PBD model, for a particular choice of the  $P_{AT}$
and $E_n$ parameters, is predominantly characterized by a single mLE value $\chi_1$, except for some extreme cases
discussed in Sect.~\ref{ssec:alpha} below. A reasonable estimation of this value  can be obtained by the average
$\langle \chi \rangle$ of the final $\chi$ values, while the corresponding uncertainty is simply taken as the standard deviation
of these values.

%%%%%%%%%%%%%%%%%%%%%%%%%%%%%%%%
\begin{figure}
    \includegraphics[width=0.45\textwidth]{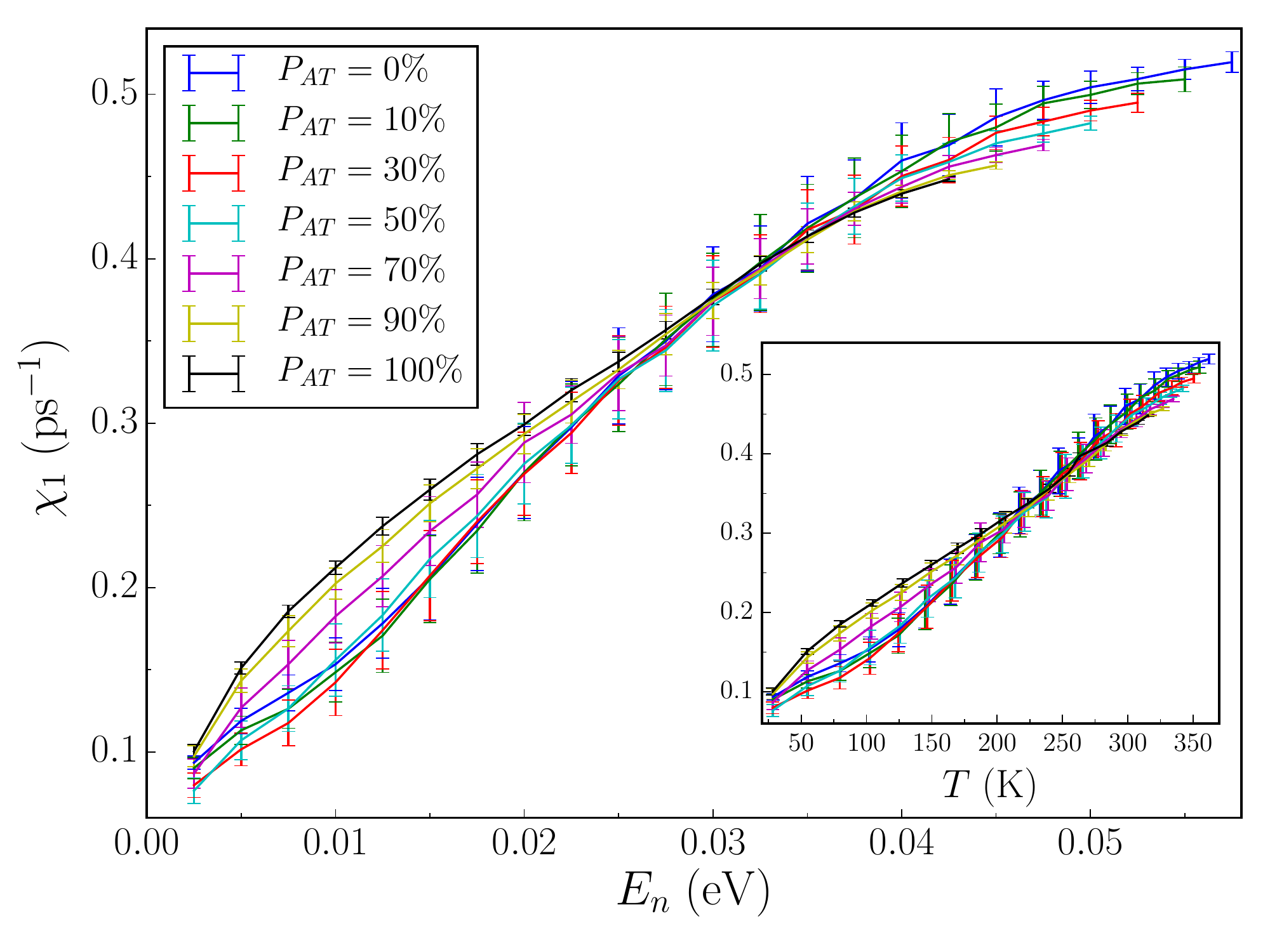}
\caption{\label{fig:allmles}(\textit{Color online}) Estimates of the system's mLE $\chi_1$ as a function of the energy density $E_n$,
for several percentages $P_{AT}$ of AT base pairs in the DNA chain, plotted up to the melting point.
The error bars represent the standard deviation in the
computation of the limiting values of $\langle \chi \rangle$. Computed points are connected with lines in order to facilitate the
visualization of the dependence of $\chi_1$ on $E_n$. \textit{Inset}: The dependence of $\chi_1$ on temperature $T$.}
\end{figure}
%%%%%%%%%%%%%%%%%%%%%%%%%%%%%%%%

Repeating the procedure of Fig.~\ref{fig:mlebuild} for several values of $P_{AT}$ and $E_n$ we can find the behavior of the system's mLE as the energy density and the content of AT base pairs change. The results of this process are presented in Fig.~\ref{fig:allmles} where each curve corresponds to a particular $P_{AT}$ value. The estimates of $\chi_1$  are calculated up to the chain's melting point for each $P_{AT}$. According to figure 2(b) of Ref.~\cite{saul2} the melting temperature $T_m$ (in K) is related to  $P_{AT}$ through the equation
\begin{equation}\label{eq:T_m}
    T_m = 365 - 0.4P_{AT}.
\end{equation}
Note that the slope of 0.4 degrees per 1$\%$ percentage of AT or GC, indicated in Eq.~(\ref{eq:T_m}), is in agreement with
experimental measurements over a large number of different DNA samples \cite{MD62}. In our simulations the temperature $T$ is estimated
as $T=2 \overline{E_n^{kin}}/k_B$, where $\overline{E_n^{kin}}=\left( \sum_{i=1}^n p_i^2/2m \right)/ n$ is the mean kinetic energy
per base pair and $k_B$ is the Boltzmann constant. In Fig.~\ref{fig:allmles} we present for each AT concentration
results up to $E_n$ values  ending at the denaturation transition defined in Eq.~(\ref{eq:T_m}).
The melting of chains having more GC pairs (with the extreme situation being the $P_{AT}=0\%$ case) happens at larger
$E_n$ values. This is due to the fact that a GC base pair contains three hydrogen bonds, in contrast to AT pairs which have only
two such bonds, and consequently more energy is needed for GC bonds  to break. This situation is reflected in the parameters of the model.
For this reason curves of smaller $P_{AT}$  extend to higher $E_n$ values in Fig.~\ref{fig:allmles}.

For all percentages, the same trend is clearly visible in Fig.~\ref{fig:allmles}: the $\chi_1$ values (and consequently the system's
chaoticity) increases  with energy density, with a minor flattening out at larger $E_n$ values. This behavior is in agreement
with the results presented in \cite{barreEPL} for the homogeneous {PBD model, but for different
parameter values from the ones used here. From Fig.~\ref{fig:allmles} we observe  slightly different behaviors at lower and
higher energies for different values of $P_{AT}$. At lower energy densities, $E_n \lesssim 0.025$ eV, there is a tendency
for the chains with higher AT content to be more chaotic, as indicated by the higher values of the estimated mLEs. In this energy
region it is evident that the curve of the homogenous AT chain ($P_{AT}=100\%$) is always at higher values than all other cases.
However, in this energy range the $P_{AT}=0\%$ case is not always the least chaotic one; there is some crossing point, at about $E_n \approx 0.015$eV, below which
the $P_{AT}=10\%$ and $P_{AT}=30\%$ cases are less chaotic than the pure GC case. In the middle region,
$0.025 \; \textrm{eV} \lesssim E_n \lesssim 0.035$ eV, the composition of the chain appears to have little effect on the system's
chaoticity. All $P_{AT}$ curves  show very similar behaviors until  $E_n \approx 0.035$ eV, whereafter the chains with more
GC content seem to become somewhat more chaotic, up to the melting point. In particular, for $E_n \gtrsim 0.035$ eV the curve
which corresponds to a pure GC chain ($P_{AT}=0\%$) is always at higher $\chi_1$ values than the other curves.
 Following Ref. \cite{T-Rpre}, we have obtained a critical exponent, through the relation
 $|\chi_1(T) - \chi_{1m}| \propto (T_m - T)^a$, where $\chi_1(T)$ is the mLE at temperature $T$ approaching melting and
 $\chi_{1m}$ is the mLE at the melting temperature $T_m$. A value of this exponent $a=1.5\pm0.2$ has been obtained for
 all different AT percentages.
 
The existence of these three different dynamical regimes in the chaotic behavior of the PBD system is  consistent with the easier breaking of AT bonds. For relatively small energies it is expected that the stronger GC bonds lead to smaller oscillations, resulting to less chaotic behavior. Thus, chains with more AT base pairs show larger $\chi_1$ values and behave more chaotically in this regime as these base pairs explore in larger extent the nonlinear part of the corresponding Morse potential. On the other hand,  larger energies (but not large enough to lead to the breaking of base pairs and to melting) correspond to higher energy levels for each base pair. These high energy levels in the narrower well of the GC Morse potential (with respect to the AT pair) lead to more chaos for the GC pairs than the AT ones. In the case of AT base pairs  these high energy levels are closer to the rim of their Morse potential, allowing larger $y_n$ values, which in turn make their behavior more linear. Thus, in that energy regime lattices with small $P_{AT}$ values become more chaotic. The smooth increase of $\chi_1$ curves between these two energy regimes results in an intermediate region ($0.025 \; \textrm{eV} \lesssim E_n \lesssim 0.035$ eV) where the value of $P_{AT}$ practically does not influence the system's mLE.
It is also expected that for very small $E_n$ values (i.e.~$E_n\rightarrow0$ eV) the nonlinear effects in Eq.~(\ref{eq:hamiltonian}) should become negligible and the system will become less chaotic, irrespectively of its content of GC and AT base pairs. This tendency is present in Fig.~\ref{fig:allmles} as the $\chi_1$ estimates are reduced when $E_n$ becomes very small, attaining values, for all $P_{AT}$, which are closer to each other.

In order to examine whether biologically functional sequences exhibit characteristic chaotic properties,
we have investigated two particular gene promoters: a 86 base-pair long segment of the adenovirus major late promoter (AdMLP)
and a 129 base-pair long fragment of \textit{lac} operon promoter. The base-pair sequences of these DNA stretches are shown
in Refs. \cite{EPL} and \cite{angeliki}, respectively. The statistically averaged mLEs of these sequences are identical with the
corresponding values obtained from random sequences of the same length and AT percentage ($P_{AT}=33.7\%$ and 51.9\%
for the  AdMLP and \textit{lac} operon, respectively), thus indicating no specific chaoticity of gene promoters.

\begin{figure}
    \includegraphics[width=0.45\textwidth]{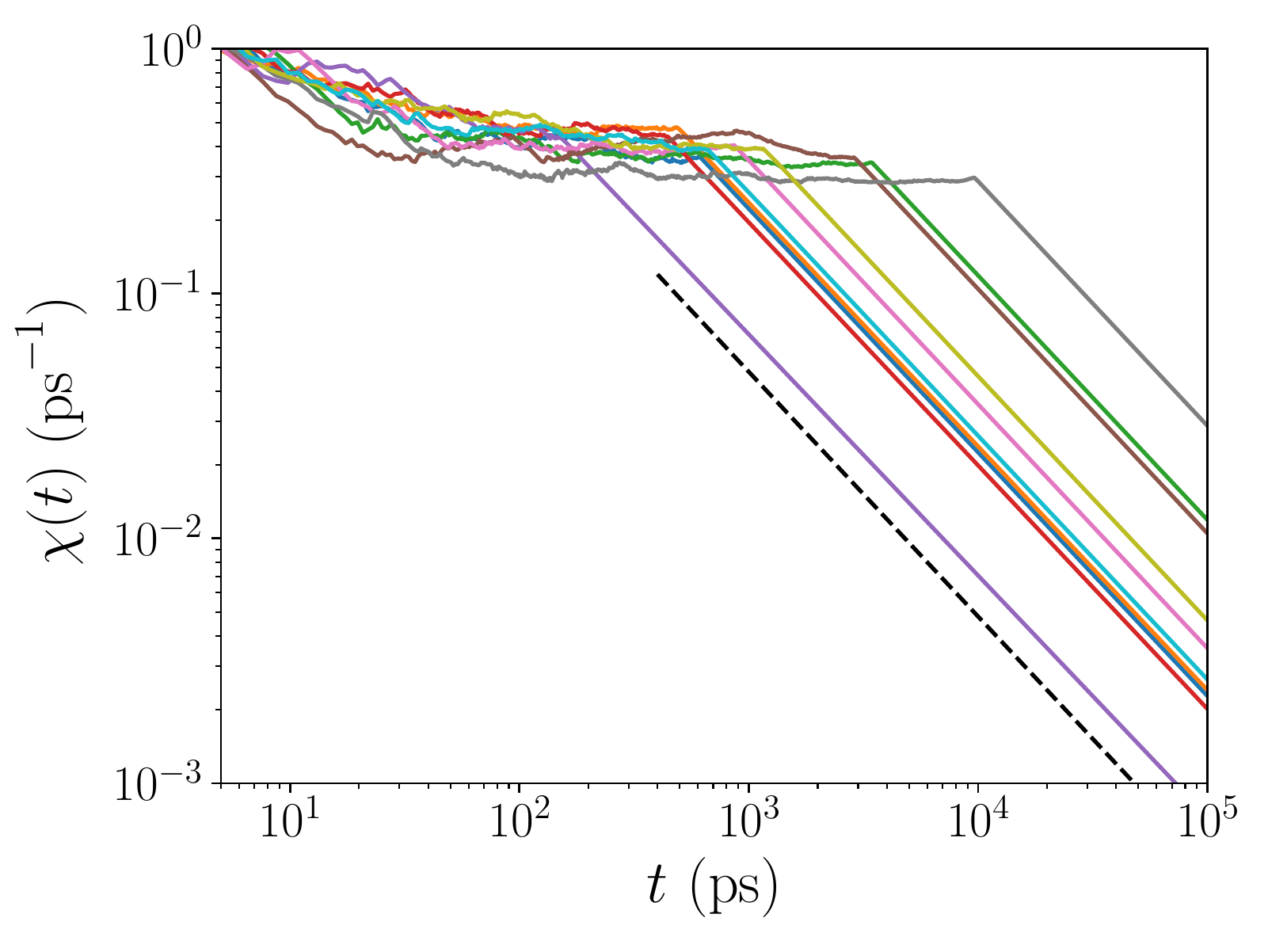}
    \caption{\label{fig:postmelting}(\textit{Color online}) Evolutions of finite time mLEs $\chi$ above the melting temperature,
for ten different initial conditions of a single disorder realization for the case of $P_{AT}=90\%$ at $E_n=0.085$ eV.
Once the chain melts and the bonds completely separate, the system linearises and the finite time mLE $\chi$ shows
a behavior corresponding to a regular orbit. The dashed line guides the eye at a slope of $-1$ in the log-log scale.}
\end{figure}

Above melting the base pair stretchings $y_n$ are boundless in the framework of the PBD model and the exponential
terms in the Hamiltonian of Eq.(\ref{eq:hamiltonian}) drop off to zero, leading to effectively integrable behavior.
The evolution of the finite time mLE in this case is shown in Fig.~\ref{fig:postmelting}. It can be seen the after some time $\chi$
starts going to zero with a slope of $-1$ in a log-log scale, as is expected for a regular orbit (see for example \cite{S_10} and references therein).

\subsection{Deviation Vector Distributions}\label{ssec:DVD}

Based on the fact that deviation vectors eventually are aligned to the direction defined by the mLE, DVDs have already been used to visualize the motion of chaotic seeds, i.e.~regions of high $\xi_i$ (Eq.~\ref{eq:dvd}) values, in chaotic, nonlinear lattices \cite{skokosdvd,SMS18}. In this section we examine the spatiotemporal evolution of such DVDs in conjunction with the
displacements of the system.

Looking at the time evolution of individual DVDs we can see a correlation between regions of relatively large
displacements and the concentration of the DVD. This correlation has been observed for different AT percentages and
various energy densities that we have examined. A typical result is shown in Fig.~\ref{fig:dvddisp}. The DVD is always
quite localized and appears to jump, with no apparent pattern, between sites next to a relatively large
displacement. However, it is always localized close to sites exhibiting base pair openings.
From this we can infer that in sites nearby to relatively larger base pair stretchings in the DNA chain the behavior is particularly
nonlinear, in the sense that the DVD is concentrated in these regions. There seems however to be no particular favor
shown to the size of the openings, with the DVD at different times localized around relatively larger or smaller openings
and not in the highest opening. It does however appear to completely avoid non-excited regions.
%%%%%%%%%%%%%
\begin{figure}
    \includegraphics[width=0.5\textwidth]{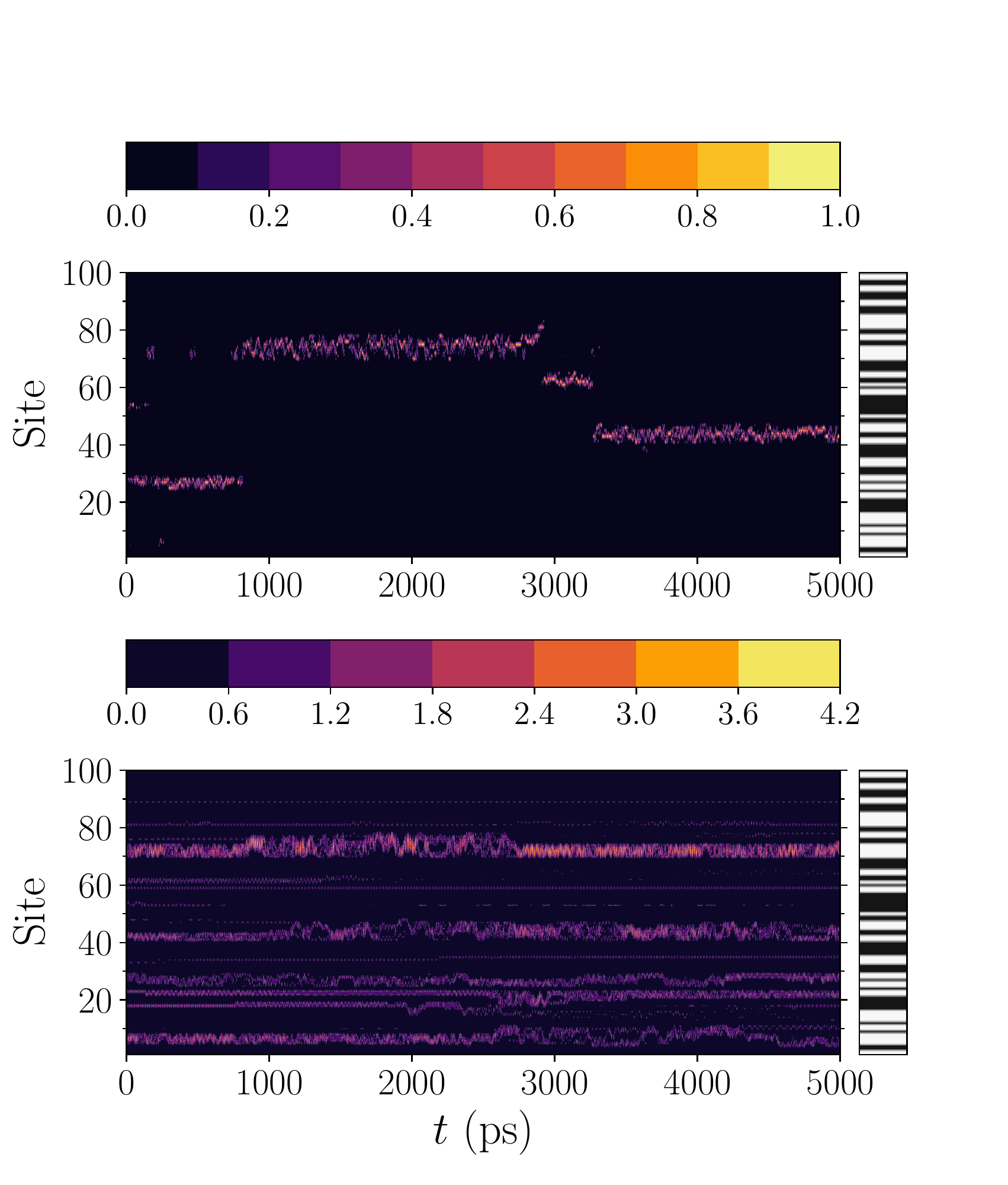}
\caption{\label{fig:dvddisp} (\textit{Color online}) ({\it Top}): Evolution of the DVD in the case of one initial condition for a particular disorder
realization. The light regions are where the DVD is concentrated, according to the color bar on the top of the figure.
({\it Bottom}): Evolution of base pair displacements for the same realization. Light colors signify high values
and dark no significant displacement according to the color bar on the top of the figure (values are shown in \AA).
The bars on the right hand side indicate the DNA sequence, showing the positions of AT (white) and GC (black) base pairs.
Energy density is
$E_n = 0.04$ eV and $P_{AT}=50\%$.}
\end{figure}
%%%%%%%%%%%%%

At low energies, looking at the averaged DVD and displacement patterns, over several initial conditions, for a particular
disorder realization, one can clearly distinguish the influence of the base pair distribution along the DNA sequence.
The displacements are larger on AT base pairs and especially in AT-rich regions, as expected.
The DVD tends to be concentrated in the larger homogeneous islands, i.e. in regions containing the larger number
of consecutive sites of the same type of base pair.
This is demonstrated in Fig.~\ref{fig:dvdavg}, where the averaged DVD and displacement patterns are shown for a hundred
different initial conditions of energy density $E_n = 0.005$ eV, corresponding to the same disorder realization with $P_{AT}=30\%$.
While the displacements are larger at the AT sites, the DVD has a clear trend towards being concentrated at the larger
homogeneous islands, regardless of the base pair type of the island. This is a typical situation for any AT percentage
at lower energies. However at higher energies (e.g. at $E_n = 0.03$ eV), while the displacements are always larger
at the AT base pairs, the averaged DVD loses its coherence with the larger homogeneous islands
and is no longer concentrated there.
%%%%%%%%%%%%%
\begin{figure}
    \includegraphics[width=0.5\textwidth]{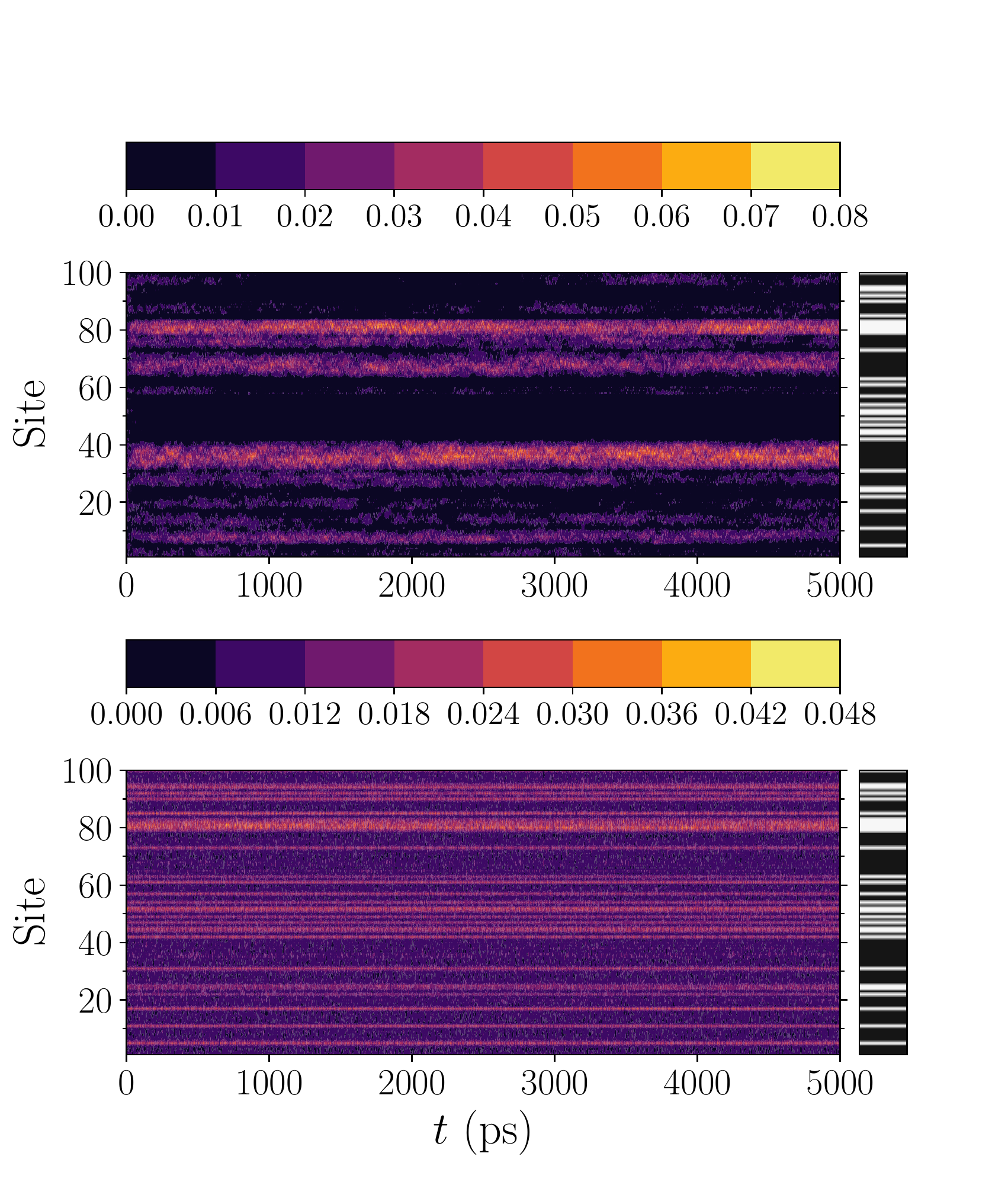}
\caption{\label{fig:dvdavg} (\textit{Color online}) ({\it Top}): Evolution of the DVD 
at low energy density, averaged over 100 initial conditions for a single disorder realization. The light regions are where the DVD is
concentrated, according to the color bar on the top of the figure.
({\it Bottom}): Evolution of base pairs displacements averaged over the same 100 initial conditions. Light colors signify maximum
values and dark no significant displacement, according to the color bar on the top of the figure (values are shown in \AA).
The bars on the right hand side indicate the DNA sequence, showing the positions of AT (white) and GC (black) base pairs.
$E_n = 0.005$ eV, $P_{AT}=30\%$. }
\end{figure}
%%%%%%%%%%%%%

Using long simulations (up to $10^5$ ps), we have also computed the evolution of the DVD, averaged over 100
different realizations, for the two functional sequences considered previously, the AdMLP and \textit{lac} operon promoters,
at energies corresponding to a physiological temperature. The results are presented in Fig.~\ref{fig:dvdadmlplac}.
As it can be seen from this figure (top), the DVD for the AdMLP case distinctly avoids a region starting close to
the transcription start site $+1$ and extending upstream up to around $-20$. This is a region presenting the lower propensity
to form large bubbles in the considered promoter stretch \cite{raptiBJ}. The DVD evolution for the \textit{lac} operon
(Fig.~\ref{fig:dvdadmlplac}, bottom) does not exhibit such a clear avoidance at any region and no correlation with the
known bubble opening probabilities \cite{angeliki} can be made from these data.
%%%%%%%%%%%%%%%%%%%%%
\begin{figure}
    \centering
    \includegraphics[width=0.5\textwidth]{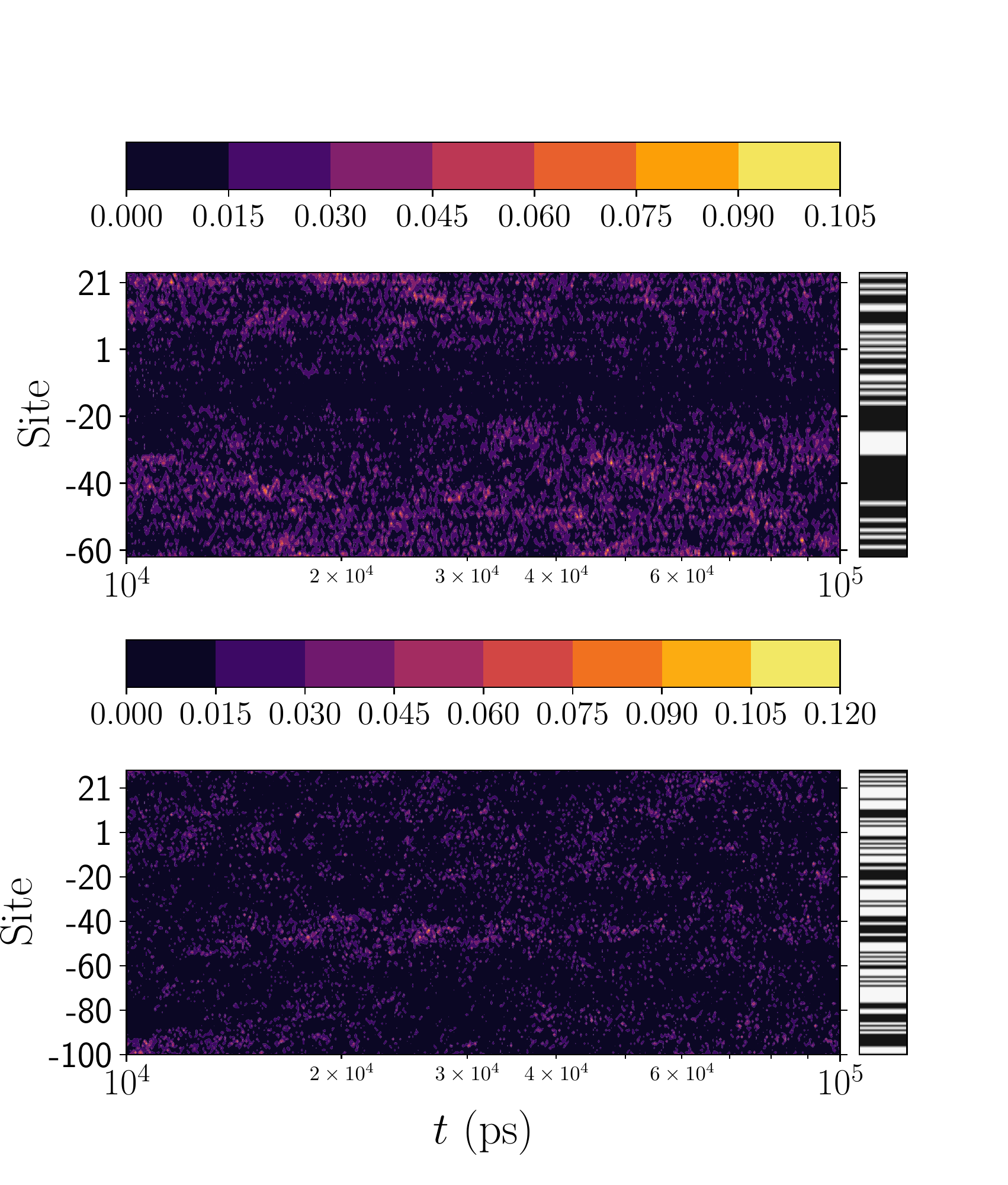}
    \caption{\label{fig:dvdadmlplac} (\textit{Color online}) ({\it Top}): Evolution of the DVD for the 86 base-pair long
AdMLP promoter, averaged over 100 initial conditions at $E_n = 0.04$ eV, corresponding to $T\approx310$ K.
({\it Bottom}): Evolution of the DVD for the 129 base-pair long \textit{lac} operon, averaged over 100 initial conditions at the same
energy density. The light regions show where the DVDs are concentrated, according to the color bars on the top of each figure.
The bars on the right hand side indicate the DNA sequence, showing the positions of AT (white) and GC (black) base pairs for the two sequences.}
\end{figure}
%%%%%%%%%%%%%%%%

\subsection{Effect of the alternation index\label{ssec:alpha} }

In order to further investigate the effect of heterogeneity on the chaoticity of the system, following the influence of the
AT/GC composition, we now examine how the estimate of $\chi_1$ changes with the alternation index $\alpha$.
In \cite{RCDalpha} it was shown that the probability distribution function of $\alpha$ can be well approximated by a rather narrow Gaussian distribution. Thus, in our analysis we consider the most probable value of $\alpha$, along with the extreme cases of very large $\alpha$, corresponding
to a well-mixed disorder realization which can be thought to be effectively homogeneous, as well as the case of very small $\alpha$,
corresponding to a very unevenly distributed or heterogeneous disorder realization.
Particularly, we take the smallest possible value of $\alpha$ ($\alpha=2$, where the AT and GC base pairs are completely
separated in two distinct parts along the DNA chain), another small value $\alpha=6$, and similarly the maximum value of $\alpha$
(where the minority base pairs, AT or GC, are all isolated in between base pairs of the other type, the majority one) along with another large $\alpha$ value. We note that the maximum possible value of $\alpha$ depends on  $P_{AT}$  \cite{RCDalpha}.

\begin{figure}
    \centering
    \includegraphics[width=0.48\textwidth]{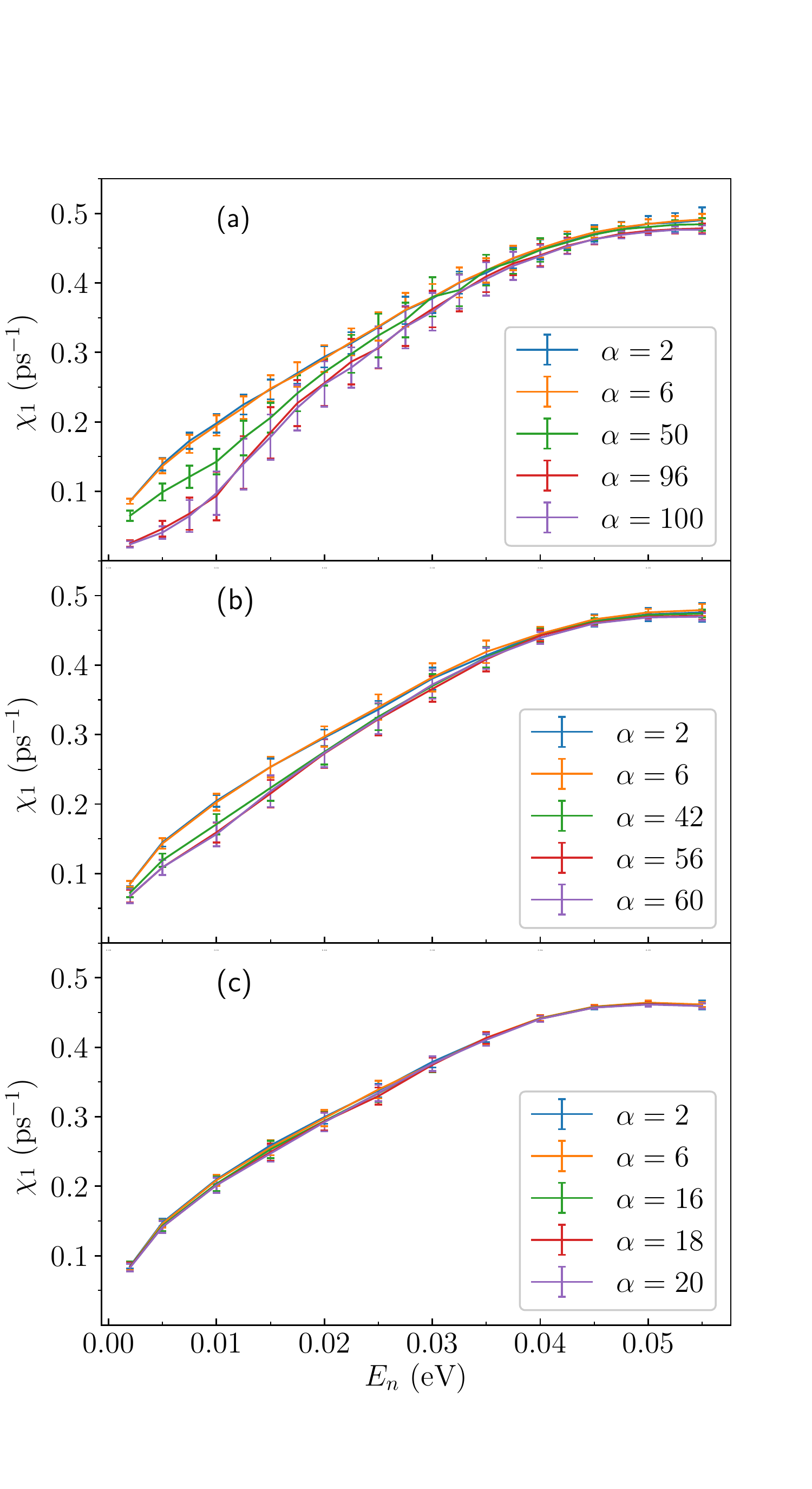}
    \caption{\label{fig:alphamle} (\textit{Color online}) Estimate of the mLE $\chi_1$  as a function of the energy density $E_n$ for
different values of the alternation index $\alpha$, for AT percentages (a) $P_{AT}=50\%$, (b) $P_{AT}=70\%$ and (c) $P_{AT}=90\%$.
In (a) the extremely heterogeneous cases ($\alpha =2, 6$) are more chaotic at lower energies than either the most probable
case ($\alpha=50$), or the effectively homogeneous cases ($\alpha =96,100$). This effect is less noticeable in (b) and it is not
observed in (c).}
\end{figure}

The dependence of the limiting value of the finite time mLE on the energy density for all these  cases of the alternation index is depicted in Fig.~\ref{fig:alphamle}(a), when the AT percentage is $P_{AT}=50\%$.
The effects of $\alpha$ can be clearly seen at low energies, below $E_n \approx 0.03$ eV: The effectively  homogeneous cases
($\alpha$=100 and 96) show lower chaoticity than the most probable case ($\alpha$=50), while the more heterogeneous cases
($\alpha$=2 and 6) are noticeably more chaotic. As the energy is increased to the melting point, the effect of $\alpha$ becomes negligible, but the overall picture of the homogeneous cases being less chaotic remains slightly apparent.

We find that unsurprisingly the effect of $\alpha$ on the system is most noticeable in the $P_{AT}=50\%$  case, where there is
an equal number of the two different types of base pairs. In systems dominated by a single base pair type, the effect
of the few other base pairs is minimal regardless of the disorder arrangement. This can be seen in Figs. \ref{fig:alphamle}(b)
and \ref{fig:alphamle}(c), demonstrating similar results as in Fig.~\ref{fig:alphamle}(a) for $P_{AT}=70\%$ and $90\%$,
respectively. As compared to the $P_{AT}=50\%$ case, the effect of the value of the alternation index $\alpha$ is smaller
at $P_{AT}=70\%$  and almost vanishes at $P_{AT}=90\%$. Similar results to $P_{AT}=70\%$ and $P_{AT}=90\%$
have been obtained when $P_{AT}=30\%$ and $P_{AT}=10\%$, respectively.

%_______________________________________________
\section{Conclusions\label{sec:sum}}

We have calculated the finite time maximum Lyapunov exponent $\chi$ of the Peyrard-Bishop-Dauxois DNA model,
for different base pair compositions across the whole energy spectrum up to the melting point. Chaoticity increases
with energy for any composition of AT/GC base pairs, including the homogeneous cases of pure AT or pure GC chains.
Three distinct regions of chaotic behavior have been found: (i) at lower energy densities ($E_n \lesssim 0.025$ eV), DNA segments
with more AT base pairs are more chaotic, (ii) in a middle energy region the composition of the chain has little impact on the chaoticity,
and (iii) at higher energy densities ($E_n \gtrsim 0.035$ eV) sequences with more GC base pairs appear to be slightly more chaotic.
The maximum Lyapunov exponent of biologically functional gene promoters does not exhibit a distinct behavior compared
to random sequences of the same AT percentage.

The deviation vector distribution, which can identify regions of more intense chaotic behavior,  has been found in individual
realizations to be localized in regions along the DNA chain
close to relatively large displacements. The numerical simulations show that the DVD jumps between such regions nearby to
a large displacement, with no particular preference to the sites of highest displacement. In addition, for relatively low energies
a tendency of the statistically averaged DVDs to be localized in extended regions containing the same type of base pairs was found.

Another measure of heterogeneity, the alternation index $\alpha$, appears to have some effect on the maximum Lyapunov
exponent in cases where there is not a predominance of one type of base pair, especially at low energies. In particular, our findings show that in these cases effectively homogeneous segments (large values of $\alpha$) are generally less chaotic than more heterogeneous segments (where $\alpha$ is small).

%_______________________________________________
\begin{acknowledgments}
Ch.S.~and G.K.~were supported by the Erasmus+/ International Credit Mobility KA107 program. Ch.S.~acknowledges support by the National Research Foundation (NRF)
of South Africa (IFRR and CPRR Programmes). M.H.~acknowledges financial assistance from the NRF. The authors acknowledge the Centre for High Performance Computing (\url{https://chpc.ac.za}) in South Africa, for providing computational resources to this research project.
\end{acknowledgments}

%_______________________________________________


\begin{thebibliography}{99}

\bibitem{physrep} M. Manghi and N. Destainville, Phys. Rep. {\bf631}, 1 (2016).

\bibitem{PBD} T. Dauxois, M. Peyrard, and A.R. Bishop, Phys. Rev. E {\bf 47}, R44 (1993).

\bibitem{CG} A. Campa and A. Giansanti, Phys. Rev. E {\bf 58}, 3585 (1998).

\bibitem{boianNAR9} B.S. Alexandrov, V. Gelev, Y. Monisova, L.B. Alexandrov, A.R. Bishop, K.\O. Rasmussen, and
A. Usheva, Nucleic Acids Res. {\bf 37}, 2405 (2009).

\bibitem{CH} D. Cule and T. Hwa, Phys. Rev. Lett. {\bf 79}, 2375 (1997).

\bibitem{aresPRL} S. Ares, N.K. Voulgarakis, K.\O. Rasmussen, and A.R. Bishop, Phys. Rev. Lett. {\bf 94}, 035504 (2005).

\bibitem{NAR} C.H. Choi, G. Kalosakas, K.\O. Rasmussen, M. Hiromura, A. Bishop, and A. Usheva,
Nucleic Acids Res. {\bf 32}, 1584 (2004).

\bibitem{EPL} G. Kalosakas, K.\O. Rasmussen, A.R. Bishop, C.H. Choi, and A. Usheva,
Europhys. Lett. {\bf 68}, 127 (2004).

\bibitem{raptiBJ} C.H. Choi, Z. Rapti, V. Gelev, M.R. Hacker, B. Alexandrov, E.J. Park, J.S. Park, N. Horikoshi,
A. Smerzi, K.\O. Rasmussen, A.R. Bishop, and A. Usheva, Biophys. J. {\bf 95}, 597 (2008).

\bibitem{boianPLOSCB} B.S. Alexandrov, V. Gelev, S.W. Yoo, A.R. Bishop, K.\O. Rasmussen, and A. Usheva,
PLoS Comput. Biol. {\bf 5}, e1000313 (2009).

\bibitem{boianNAR2} B.S. Alexandrov, V. Gelev, S.W. Yoo, L.B. Alexandrov, Y. Fukuyo, A.R. Bishop,
K.\O. Rasmussen, and A. Usheva, Nucleic Acids Res. {\bf 38}, 1790 (2010).

\bibitem{angeliki} A. Apostolaki and G. Kalosakas, Phys. Biol. {\bf 8}, 026006 (2011).

\bibitem{faloPRE12}  R. Tapia-Rojo, D. Prada-Gracia, J.J. Mazo, and F. Falo, Rhys. Rev. E {\bf 86}, 021908 (2012).

\bibitem{huangJBE} H.-H. Huang and P. Lindblad, J. Biol. Eng. {\bf 7}, 10 (2013).

\bibitem{faloPLOS}  R. Tapia-Rojo, J.J. Mazo, J.A. Hernandez, M.L. Peleato, M.F. Fillat and F. Falo,
PLoS Comput. Biol. {\bf 10}, e1003835 (2014).

\bibitem{PB} M. Peyrard and A.R. Bishop, Phys. Rev. Lett. {\bf 62}, 2755 (1989).

\bibitem{DPB93} T. Dauxois, M. Peyrard, and A.R. Bishop, Phys. Rev. E {\bf 47}, 684 (1993).

\bibitem{peyrardRev} M. Peyrard, Nonlinearity {\bf 17}, 1 (2004).

\bibitem{DP95} T. Dauxois and M. Peyrard, Phys. Rev. E {\bf 51}, 4027 (1995).

\bibitem{saul1} S. Ares and G. Kalosakas, Nano Lett. {\bf 7}, 307 (2007).

\bibitem{saul2} G. Kalosakas and S. Ares, J. Chem. Phys. {\bf 130}, 235104 (2009).

\bibitem{theodPRE} N. Theodorakopoulos, Phys. Rev. E {\bf 77}, 031919 (2008).

\bibitem{VKRB} N.K. Voulgarakis, G. Kalosakas, K.\O. Rasmussen, and A.R. Bishop, Nano Lett. {\bf 4}, 629 (2004).

\bibitem{CPL} G. Kalosakas, K.\O. Rasmussen, and A.R. Bishop, Chem. Phys. Lett. {\bf 432}, 291 (2006).

\bibitem{voulgSR} J.J. Traverso, V.S. Manoranjan, A.R. Bishop, K.O. Rasmussen, and N.K. Voulgarakis,
Sci.Rep. {\bf 5}, 9037 (2015).

\bibitem{farago}  M. Peyrard and J. Farago, Physica A {\bf 288}, 199 (2000).

\bibitem{maniadisPRE} P. Maniadis, B.S. Alexandrov, A.R. Bishop, and K.\O. Rasmussen, Phys. Rev. E {\bf 83}, 011904 (2011).

\bibitem{JCP03} G. Kalosakas, K.\O. Rasmussen, and A.R. Bishop, J. Chem. Phys. {\bf 118}, 3731 (2003).

\bibitem{PRE05} G. Kalosakas, K.L. Ngai, and S. Flach, Phys. Rev. E {\bf 71}, 061901 (2005).

\bibitem{PRE11} G. Kalosakas, Phys. Rev. E {\bf 84}, 051905 (2011).

\bibitem{gufu} L. Gu and H.-H. Fu, New J. Phys. {\bf 18}, 053032 (2016).

\bibitem{barreEPL} J. Barre and T. Dauxois, Europhys. Lett. {\bf 55}, 164 (2001).

\bibitem{blanesrk} S. Blanes and P. Moan, Journ. Comp. App. Math., \textbf{142}, 313 (2002).

\bibitem{hairergeom} E. Hairer, C. Lubich, and G. Wanner, Geometric Numerical Integration, Vol. \textbf{31} (Springer, New York) 2002.

\bibitem{benettinetal} G. Benettin, G. Galgani, A. Giorgilli, and J.-M Strelcyn, Meccanica \textbf{15}, 21 (1980).

\bibitem{S_10} Ch. Skokos, Lect. Notes Phys. \textbf{790}, 63 (2010).

\bibitem{contopoulos} G. Contopoulos, L. Galgani, and A. Giorgilli, Phys. Rev. A, \textbf{18}, 1183 (1978).

\bibitem{skokostm} Ch. Skokos and E. Gerlach, Phys. Rev. E, \textbf{18}, 036704 (2010).

\bibitem{GS11} E. Gerlach and Ch.~Skokos, Discr. Cont. Dyn. Sys.-Supp. 2011, 475

\bibitem{GES12} E. Gerlach, S. Eggl and Ch. Skokos, Int. J. Bifurcation Chaos {\bf 22}, 1250216 (2012).

\bibitem{skokosdvd} Ch. Skokos, I. Gkolias, and S. Flach, Phys. Rev. Lett., \textbf{111}, 064101 (2013).

\bibitem{SMS18}   B.~Senyange, B.~Many Manda, and Ch.~Skokos, Phys. Rev. E \textbf{98} 052229 (2018).

\bibitem{RCDalpha} M. Hillebrand, G. Paterson-Jones, G. Kalosakas, and Ch. Skokos, Regul. Chaot. Dyn. \textbf{23}, 135 (2018).

\bibitem{MD62} J. Marmur and P. Doty, J. Mol. Biol. {\bf 5}, 109 (1962).

\bibitem{T-Rpre} R. Tapia-Rojo, J.J. Mazo, and F. Falo, Phys. Rev. E \textbf{82}, 031916 (2010).


\end{thebibliography}
\end{document}